\documentclass[10pt,letterpaper]{article}
\usepackage{opex3}

\begin{document}

\title{Time-resolved detection of  relative intensity squeezed nanosecond pulses  in a $^{87}$Rb  vapor }

\author{Imad H. Agha$^1$, Christina Giarmatzi$^1$, Quentin Glorieux$^2$, Thomas Coudreau$^2$, Philippe Grangier$^1$, and Ga\'etan Messin$^1$}

\address{1-Laboratoire Charles Fabry, Institut d'Optique, CNRS, Univ. Paris-Sud,  Campus Polytechnique, 91127 Palaiseau Cedex, France}
\address{2-Laboratoire Mat\'eriaux et Ph\'enom\`enes Quantiques, UMR 7162, Universit\'e Paris-Diderot CNRS, 10, rue A. Domon et L. Duquet, 75013 Paris, France}
\email{agha@enst.fr} 



\begin{abstract}
We present theoretical and  experimental results on the generation and detection of pulsed, relative-intensity squeezed light in a 
hot  $^{87}$Rb vapor. The intensity noise correlations between a pulsed probe  beam and its conjugate, generated through nearly-degenerate four-wave mixing in a double-lambda system, are studied numerically and measured experimentally  via time-resolved balanced detection. We predict and observe about -1 dB of time-resolved relative intensity squeezing with 50 nanosecond pulses at 1 MHz repetition rate. (-1.34 dB corrected for loss). \end{abstract}



\section{Introduction}

Squeezed light is a valuable resource in the fields of continuous-variable quantum information and quantum optics. These quantum states have been used as a resource in modern quantum cryptography protocols, in entanglement generation, and in universal quantum computation (\cite{book1,cerf} and references therein). Moreover, nonclassical states of light are highly desirable for applications in atom-based quantum memories \cite{honda}, as preserving the quantum properties of the state of light is the signature of such a storage system. A quantum memory, such as that based on the long-lived coherence of an atomic system \cite{appel}, is a vital component for the implementation of a quantum repeater, which allows for the extension of the range of quantum communication networks \cite{zoller}.

These quantum states of  light have been a subject of intense research since  the mid 1980s, starting with the pioneering experiment by Slusher \textit{et al.} \cite{slusher85}.  While that experiment was performed in a cavity geometry, later experiments \cite{kumar1,kumar2, raymer1} succeeded in generating squeezed light in a single pass geometry through the process of parametric amplification of ultra-short  pico and femto-second pulses in nonlinear $\chi^{(2)}$ crystals. Later, parametric processes in optical fibers were also employed successfully in generating squeezed light in an optical fiber via its $\chi^{(3)}$ nonlinearity \cite{haus}. More recently, the  $\chi^{(3)}$ optical nonlinearity of  cold atomic ensembles was employed as a  mean of generating squeezed light \cite{lambrecht96,josse} under continuous-wave excitation.

The use of pulsed excitation in the generation of squeezed light carries several advantages over continuous-wave excitation. There is a potential for pulse-shaping \cite{book2} (e.g. via phase modulators, filters, gratings) that might prove useful for interactions with atomic and molecular systems. Moreover, recent experiments have demonstrated that pulsed squeezed light can be  de-Gaussified (via projective measurements) to produce  exotic nonclassical states of light \cite{alexei} (e.g. cat states) that allow for fundamental tests of quantum mechanics as well as continuous-variable quantum computing. One additional benefit of generating pulsed squeezed light is that it can be more easily stored and retrieved from atom-based memories \cite{honda,appel}, hence demonstrating their capacity as quantum storage systems. Finally, let us note that the absence of an optical cavity (usually present in continuous wave experiments) allows for the generation of spatially multimode quantum states involved, for example, in quantum imaging \cite{QuantumImaging,Boyer08}.

Until recently,  the generation of squeezed vacuum in a single pass geometry has mainly relied on ultra-short pulses in off-resonant systems ( $\chi^{(2)}$ crystals and optical fibers) which renders the bandwidth  incompatible with the narrow linewidths of  atomic and ionic systems. As such systems allow for deterministic quantum optical operations as well as serving as quantum memories, it is highly desirable to produce squeezed light over a narrow bandwidth as well as close to an atomic resonance. One option is to produce the squeezed light in the atomic system itself. Recently, an experiment by McCormick {\it et al.} \cite{lett1} showed that squeezed light can be produced in an atomic system, relying on the process of off-resonance parametric four-wave mixing in a rubidium   vapour.  That experiment employed continuous-wave excitation and frequency-domain detection (though later a slow-light experiment in the same group demonstrated the viability of pulsed excitation \cite{lett3} in the classical regime). While the bandwidth was compatible with atomic-based memories (a few MHz),  continuous-wave squeezing does not lend itself inherently to storage and retrieval from quantum memories. In  particular, continuous-wave  excitation necessitates  the use of optical cavities in order to define temporal modes in the  process of de-Gaussification of squeezed states of light which renders the process more complicated than the case of pulsed excitation. 

In this letter, we propose and implement a system capable of producing pulsed squeezed light based on the same principles as in Refs. \cite{lett1,coudreau2}. By employing a pulsed input, we produce  nanosecond relative-intensity squeezed pulses and employ time-resolved detection to measure the degree of squeezing obtained. This work serves as an extension of recent squeezing experiments \cite{agha}  in atomic vapors towards time-domain detection. It offers an alternative path towards quantum memory experiments requiring pulsed non-classical states of light, as well as other quantum information application that require pulsed resonant non-Gaussian states of light. 
All equations should be numbered in the order in which they appear
and should be referenced  from within the main text as Eq. (1),
Eq. (2), and so on [or as inequality (1), etc., as appropriate].

section{Basic principles}

The basic idea behind the generation of relative-intensity squeezed light as presented in this work relies on off-resonant  four-wave mixing in a double lambda-system (Fig. 1) \cite{lett1}. A pump $\omega_p$, blue-detuned from the $F=1 \rightarrow F'=2$ transition of the $^{87}Rb$ D1 line by  the one-photon detuning $\Delta_1 = \omega_p - \omega_{e,g1}$ ($ \omega_{e,g1}$ is the $F=1 \rightarrow F'=2$ excited state transition frequency), interacts with a pulsed probe beam, $\omega_s$, which is offset from the pump by approximately the hyperfine ground state separation. This offest defines the two-photon detuning $\delta =  \omega_p - \omega_{hf} - \omega_s$ , whereby $\omega_{hf}$ is the ground state hyperfine splitting.  Under suitable conditions, this energy level structure allows for the parametric amplification of the probe beam  while simultaneously creating its  conjugate through the two-photon resonance enhanced $\chi^{(3)}$ nonlinearity. The proble and conjugate beams created in four-photon scattering are correlated within the bandwidth of the process, and, quantum-mechanically, carry sub-shot noise quantum correlations \cite{lett2,coudreau}. 

\begin{figure}[htbp]
\centering \includegraphics[width=10 cm, height= 5 cm]{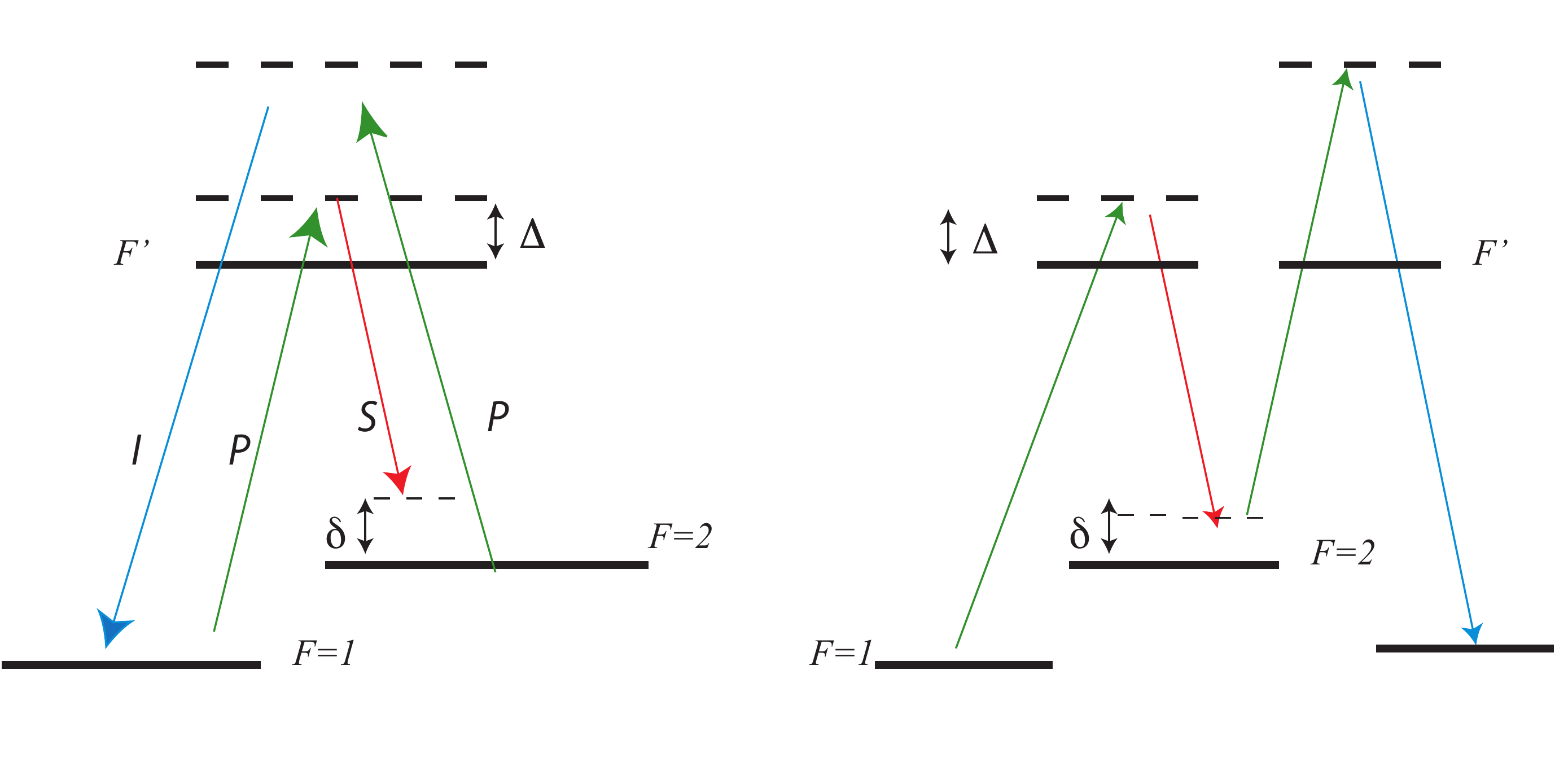}
\caption{Two \textbf{equivalent} representations of the energy-level diagram of the four-photon process. (Left) the double-lambda system showing the annihilation of two pump ($P$) photons, creating a probe ($S$) and an conjugate ($I$) photon. (Right) ``unfolded" diagram showing explicitly the two-photon resonance and the two-photon detuning. }
\end{figure}

At steady-state, and assuming no probe present, the system is optically pumped into the $F=2$ ground state via the strong pump, which allows the following recycling process to take place : $F=2 \rightarrow F=1 \rightarrow F=2$ via a double Raman transition while spontaneously creating a conjugate  and probe photons. On the other hand, optical pumping into $F=2$  allows for the process of stimulated absorption on the $F=2 \rightarrow F=1$ Raman transition. Consequently, we conclude that, in steady state, the net amplification/absorption of the  probe beam and hence the creation of the conjugate beam come from a competition between the four-wave mixing process and stimulated Raman absorption. The interplay between four-wave mixing gain and Raman absorption leads to extremely complicated dynamics that are difficult to analyze analytically. However, recent experimental results demonstrate that the four-wave mixing process can be optimized at the expense of the Raman process by adjusting the two-photon and single photon detunings of the probe and pump beams, respectively.

A numerical investigation  based upon the solution of the  Heisenberg-Langevin equations of motion for both the classical fields as well as the quantum fluctuations is presented in this section. When the parameters of the simulation are close to our experimental conditions (cell length, temperature, detector bandwidth, pump power, etc..), the results serve as a measure of  the disparity between the optimal results expected theoretically, and the experimentally obtained ones, with respect to both the classical amplification as well as quantum correlations (squeezing). 

The numerical investigation follows closely the techniques described in Ref.\cite{coudreau}. The model solves the coupled Heisenberg-Langevin and wave-evolution equations of motion for the proble/conjugate beams in the presence of a strong pump in a double-lambda system (we  note  here that the model described in \cite{coudreau}  assumes an atomic medium without velocity distribution, though it has been verified numerically that this contribution is negligible in our system). While the system was envisioned for continuous-wave excitation and frequency-resolved detection, we adapt the model to pulsed squeezed light by integrating the spectra over both the input pulse width as well as the detection bandwidth of the pulsed homodyne detectors.
The parameters of the simulation follow closely the experimental conditions to be discussed later in the paper. The simulation is run at a pump power of 750 mW, pump beam width (rad.)  of 650 $\mu m$, vapor cell temperature of 140$^\circ$C, cell length of 5 mm, and pump one-photon detuning of 1.8 GHz. 

\begin{figure}[htbp]
\centering \includegraphics[width=8 cm, height= 6 cm]{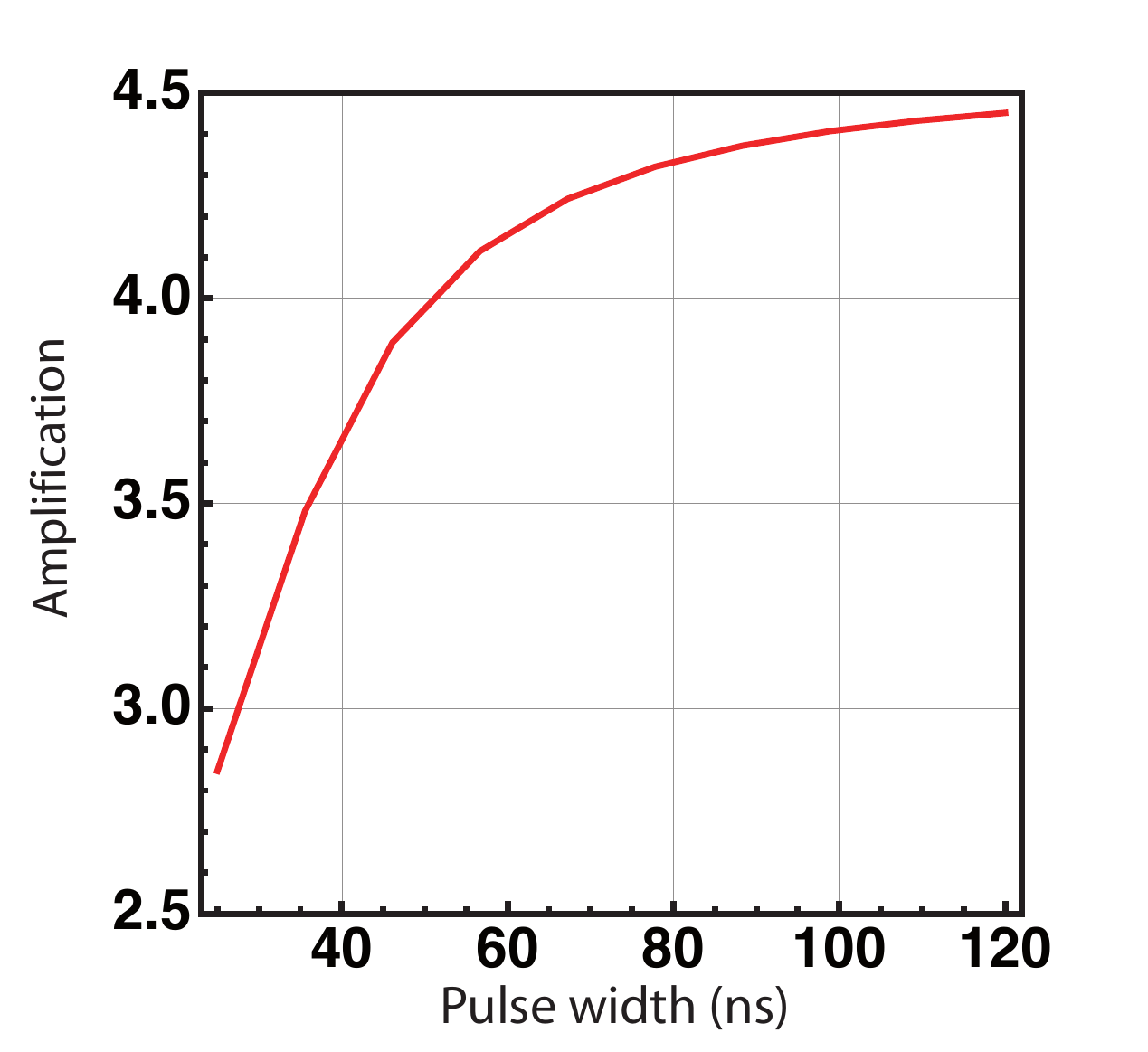}
\caption{Theoretical predictions of probe  parametric amplification vs. pulse width.}
\end{figure}
One parameter that is of interest to this experiment is the shortest pulse that can be amplified via the four-wave mixing process. Qualitatively, the shortest pulse width and consequently the largest bandwidth over which amplification takes place is given by the time interval over which the emission of the probe and conjugate photons is phase-coherent. As shown in Ref. \cite{coudreau}, the four-wave mixing gain is spectrally very  narrow (a few MHz) and the spectrum is largely dominated by stimulated Raman absorption, leading to loss on the probe and thus absence of conjugate creation. Numerically, the gain is averaged over the probe pulse spectral components to model the amplification of pulsed light (Fig. 2). The results lead us to conclude that, under our experimental conditions (mainly dictated by the laser power as well as the necessity of a wide beam to avoid transit-time broadening effects), the shortest pulse width that we can amplify is on the order of 30-50 ns.

Having theoretically demonstrated the possibility of parametrically amplifying nanosecond pulses, we next move to demonstrate the possibility of measuring  sub-shot noise quantum correlations between the amplified probe and generated conjugate pulses. As detailed in Ref. \cite{coudreau}, this is done by solving the Heisenberg-Langeving equations for the small-signal  quantum fields associated with the probe and conjugate, then calculating the noise correlations between the two quantum fields. For this calculation, we employ the parameters as above and a  detection bandwidth for the time-resolved homodyne detection of 8 MHz. Fig. 3 shows the results for narrow-band (frequency-resolved) detection, which clearly demonstrate that, for the best uniformity of squeezing vs. detection frequency, which is necessary for our time-resolved pulsed homodyne detection (which naturally integrates over the pulse spectral components and detection bandwidth), we need to set the two-photon detuning to about 20 MHz. Integrating the noise spectrum over our detection bandwidth (8 MHz), we obtain a prediction of the measurable time-resoved reduction of variance. As the two-photon detuning range for squeezing is quite narrow, \cite{coudreau}, we keep the one-photon detuning fixed at 1.8 GHz and sweep the two-photon detuning to optimize the squeezing. Fig. 4 shows the results, whereby the expected time-resolved squeezing is on the order of -1.6 dB for a 20 MHz two-photon detuning.

\begin{figure}[htbp]
\centering \includegraphics[width=8 cm, height= 6 cm]{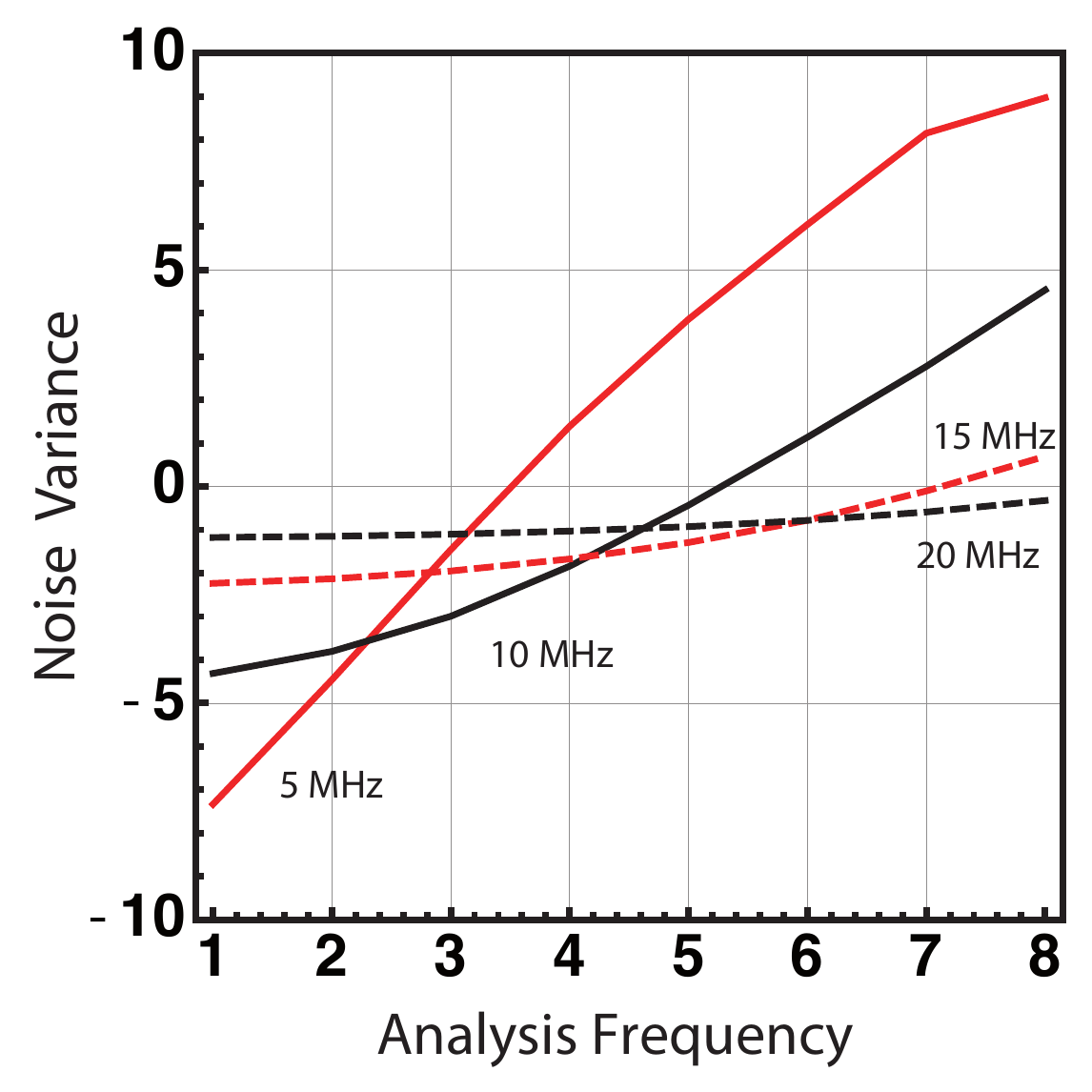}
\caption{Frequency-resolved noise power for the probe-conjugate intensity difference as a function of detection frequency for various two-photon detunings. }
\end{figure}

\begin{figure}[htbp]
\centering \includegraphics[width=8 cm, height= 6 cm]{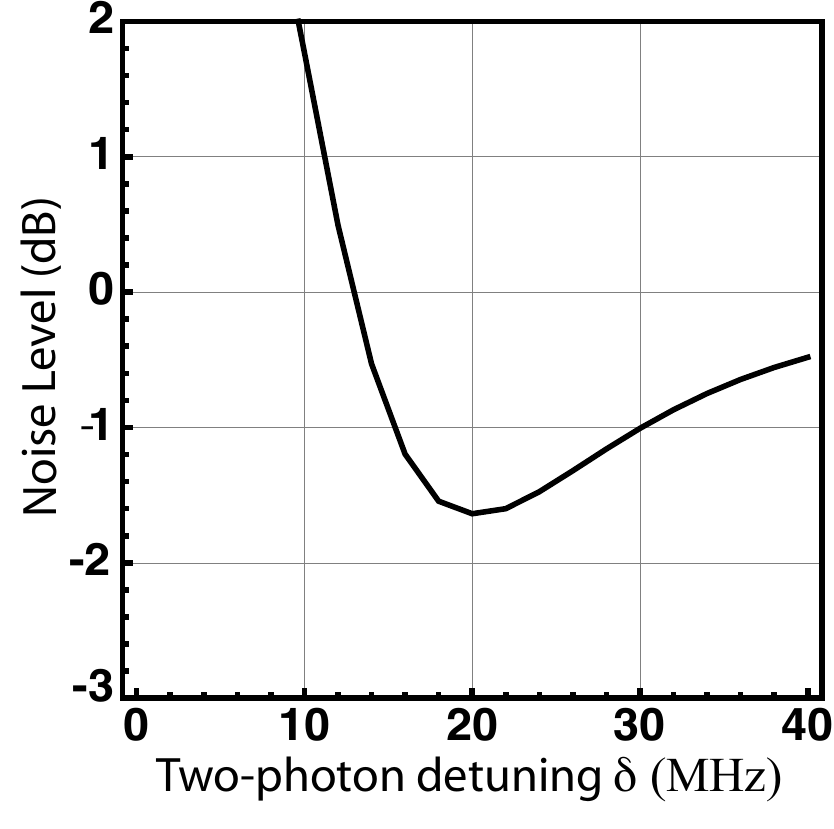}
\caption{Intensity fluctuations (time-resolved noise variance) of the probe-conjugate  photon numbers for a 50 ns pulse and an 8 MHz detection bandwidth, as a function of the two-photon detuning $\delta$. }
\end{figure}


\section{Experiment}
The experiment follows closely the procedure described in Refs.\cite{lett1, coudreau2}. A titanium-sapphire laser is blue-detuned by 1.8 GHz from the  $F=1 \rightarrow F'=2$ transition of the D1 line of $^{87}Rb$. Part of the laser beam (few mW) is coupled into an optical fiber, while the rest (750 mW) is used as a pump. The fiber-coupled beam is phase modulated via an integrated Mach-Zehnder electro-optic modulator to generate sidebands at $\pm6.83$ GHz from the pump beam frequency, which is close to the ground-state hyperfine splitting. The use of an intensity modulator to generate sidebands allows for a high ( $\approx$ 90 \%) extinction of the carrier frequency and places most of the output power in the sidebands. The excess noise at detection frequencies of interest (1-8 MHz) was measured on a spectrum analyzer showing about 1 dB of added noise, which compares favorably with side-band generation via acousto-optic modulation  \cite{lett1}. The residual carrier (pump) and one of the sidebands  are attenuated via a combination of a free-space Mach-Zehnder interferometer and  a Fabry-Perot cavity locked via the Pound-Drever-Hall  method \cite{black} and a home-built PID controller, allowing us to generate a pure tone at the probe frequency of Fig. 1. (detuned from the pump by the hyperfine splitting). The probe next passes through a free-space electro-optic modulator (LINOS 0202) producing 50 nanosecond pulses at a 1 MHz repetition rate. Experimentally, we observed that  this repetition rate minimizes the excess noise due to the intensity modulation, which was measured at about 1 to 2 dB over the detection bandwidth of a few MHz.

 The pulsed probe power  is controlled  via a  half-wave plate and a beam splitter and  is next focused down to  300 $\mu m$ inside a 0.5 cm $^{87}Rb$ vapor cell. The glass cell was placed inside a copper oven with thermo-coaxial wires wrapped around to heat the cell uniformly to 140 $^{\circ}$C. The cell and oven are then placed inside a mu-metal magnetic shield and data is only acquired outside the heating cycle (in the absence of an external magnetic field). The pump is focused to 650 $\mu m$ and overlapped with the probe (the pump beam waist is chosen to be larger than the probe to allow for a uniform amplification across the probe's beam profile)  inside the cell at a small angle ($\approx 1^{\circ}$) via a polarizing beam splitter (PBS).  The angle plays a crucial role in both the Raman (velocity selectivity) as well as the four-wave mixing (phase-matching) processes. 

\begin{figure}[htbp]
\centering \includegraphics[width=14cm, height= 11 cm]{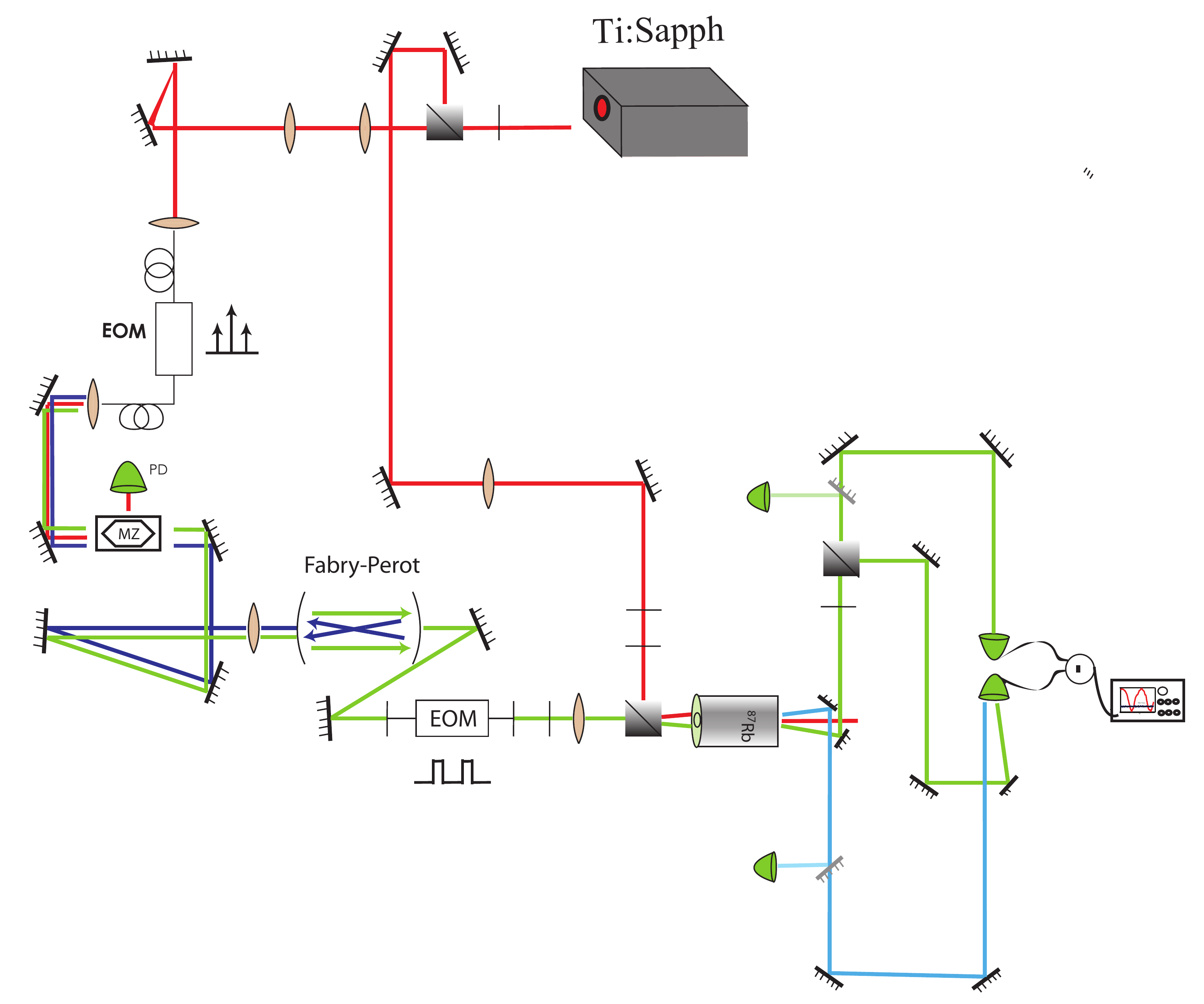}
\caption{Experimental setup for generating pulsed squeezed light in rubidium vapor.  }
\end{figure}

At the output of the cell, the pump is blocked by a polarizing beam splitter that passes the  probe  and generated conjugate. The probe and conjugate are picked off by  half mirrors while the remainder of the pump passes through and is consequently blocked. Flip-mirrors are used to alternate between classical power measurements on photodiodes  and quantum correlation measurements on a balanced detector. While the conjugate beam falls unattenuated unto the balanced detector, the probe beam is further attenuated by a $\lambda/2$ and PBS combination as to perfectly equilibrate the power on the two channels of the balanced detector, which is a requirement for  the best common-mode rejection in a homodyne detection measurement. In addition, splitting the probe beam 50:50 allows for the measurement of the shot noise quantum limit at a specific input power when the probe is far detuned from any rubidium resonances. 

\begin{figure}[htbp]
\centering \includegraphics[width= 10 cm, height=  8 cm]{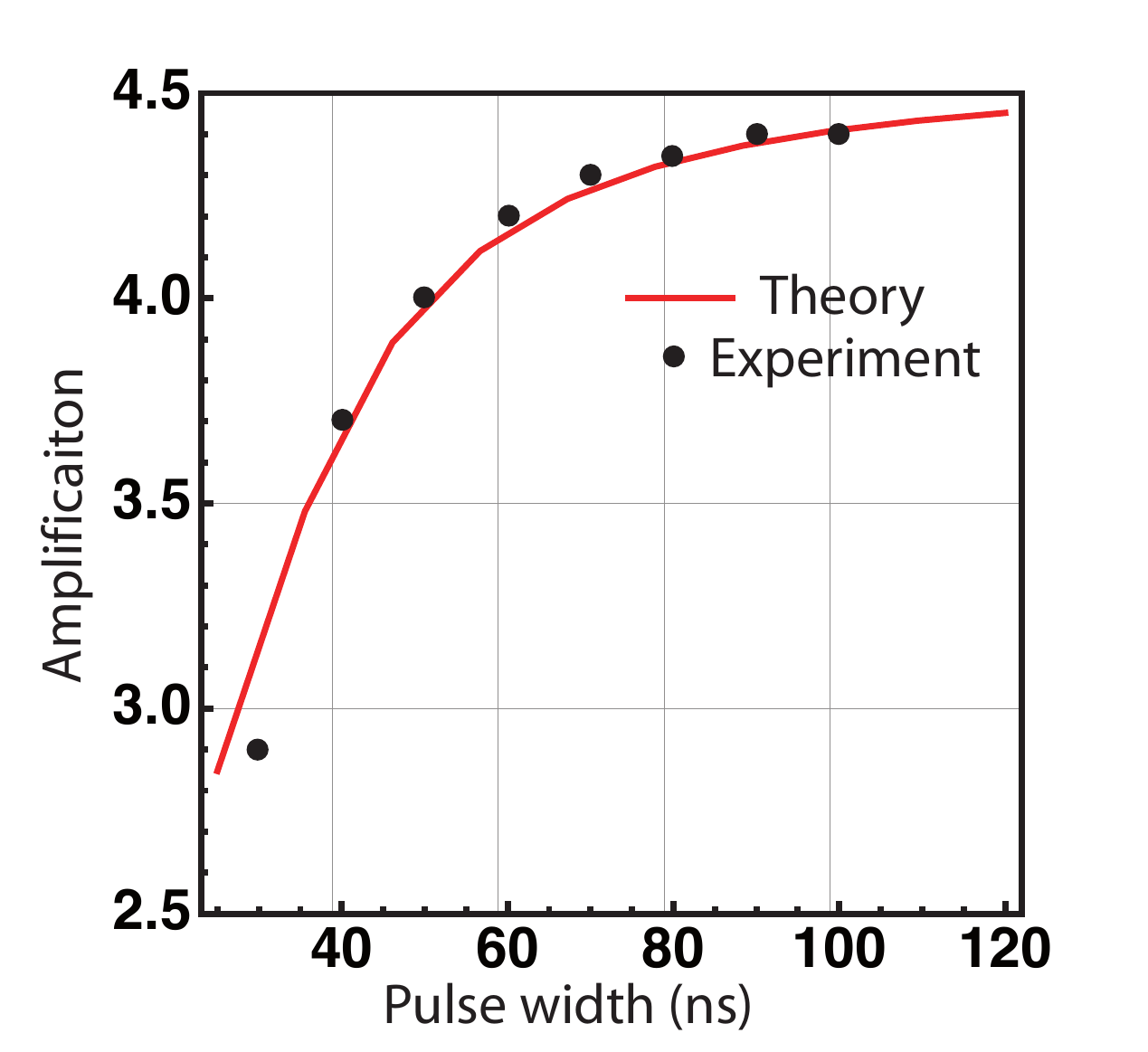}
\caption{ Gain as a function of probe pulse width. Red curve indicates theoretical prediction under our experimental conditions. Black dots indicate measured experimental values}
\end{figure}

\subsection{Classical results}
In a  four-wave mixing \hspace{0.5 mm}  $\chi^{(3)}$ parametric amplifier, the bandwidth of the amplifier and consequently the shortest pulses that can be amplified are limited by the response time of the  $\chi^{(3)}$ nonlinearity \cite{agarwal}. Considering that the coherence time between the probe and pconjugate  photons is on the order of 50 ns (Fig. 2),  we vary the pulse width between 30 and 100 ns and record the gain (at optimal parameters) on the probe. As the results of Fig. 6 show, the gain as a function of pulse width  indicates a bandwidth of a few MHz, compatible with the theoretical model given above. The agreement between the theory and experiment further confirms the validity of the theoretical model employed to predict the temporal dynamics of the four-wave mixing process. 

Hence, in our nearly-resonant system and at the optical powers we employ, we are limited by the recycling rate setting a lower limit of about 50 ns for our probe pulse beyond which we observe a decrease in the gain. With the pump detuned by 1.8 GHz from the $F=1 \rightarrow F'=2$ transition and the two-photon detuning set close to resonance, the gain on the probe pulse is measured  and plotted in Fig. 7(a), and is estimated at $4.2\pm0.1$. Simultaneously, a conjugate beam is generated in the four-wave mixing process at the opposite angle with respect to the pump, as required by phase-matching considerations. Its power is measured on a separate photodiode and the ratio of the probe/conjugate powers is plotted in Fig. 7(b). In the ideal case of no re-absorption, which is satisfied in our far-detuned case, we expect a ratio of conjugate to probe of $(G-1)/G$ \cite{lett2}. From the slope of the probe-conjugate  curve, we calculate a gain G of $4.3\pm0.2$, in agreement with the results of Fig. 7(a). This indicates that the parameters chosen favor parametric four-wave mixing at the expense of Raman absorption, as desired.

\begin{figure}[htbp]
\centering \includegraphics[width=16 cm, height=  8 cm]{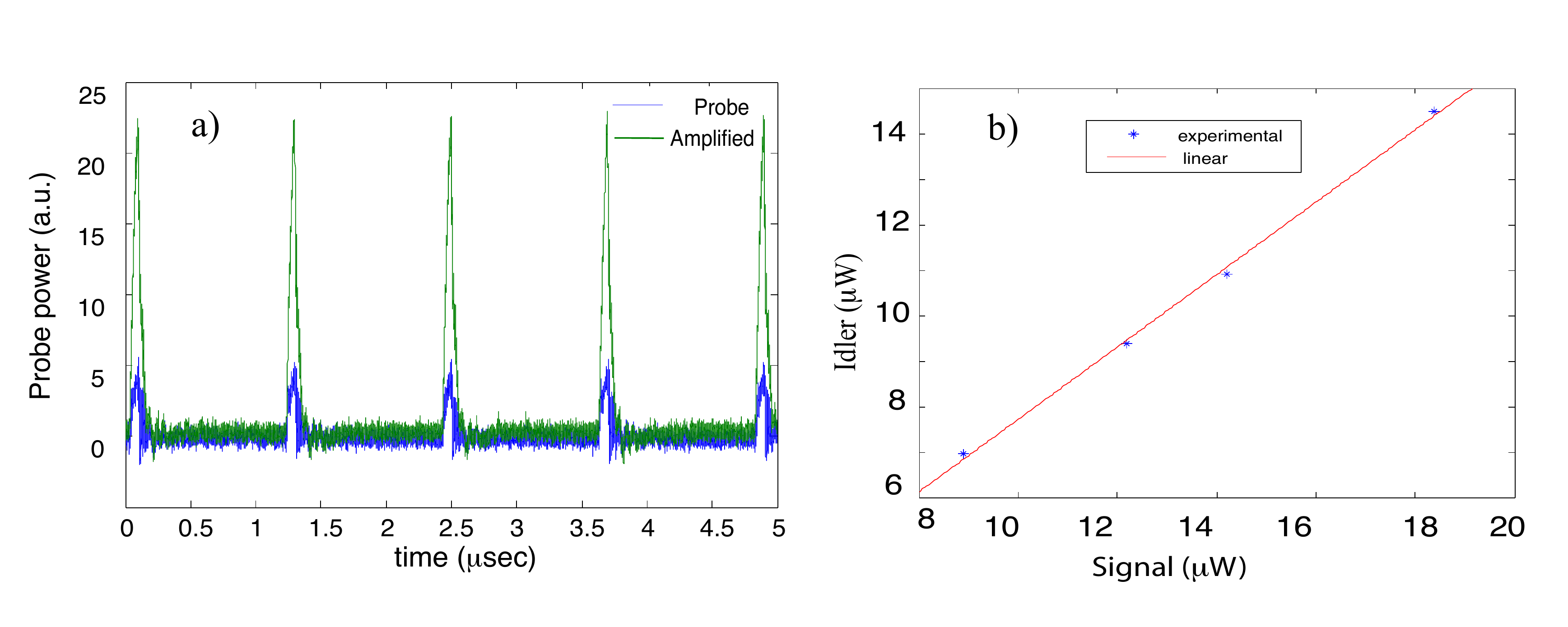}
\caption{(a) Classical gain on the probe indicating  a gain of about 4.2. (b) Produced conjugate power vs. amplified probe power. }
\end{figure}

\subsection{Quantum  measurements}
In order to measure the intensity correlations between the probe and conjugate beams and compare against the standard quantum limit, we optimize the classical gain, then we direct the probe and conjugate beams to a balanced detector in subtraction mode, which is followed by a charge-sensitive amplifier (Amtek-250) with a response time that is longer than the pulse width ($\approx 150$ ns). Data is collected over samples of 10,000 points with rolling-average subtraction to compensate for long-term drifts of the detector circuitry.  This procedure allows for the integration of the photoelectron number difference yielding an electronic pulse with a height proportional to the photon number difference between the probe and conjugate pulses. The electronic pulse  is consequently read by a data-acquisition card triggered at the same repetition rate as the EOM generating the pulsed input probe. This procedure has been proven to allow for direct estimation of the photon statistics of the input light beam \cite{raymer3} (quadrature variance in a homodyne detection, intensity correlations), which implies that what we record is a direct measure of the intensity fluctuations of the light field rather than the noise power at a certain detection frequency \cite{yurke}. The variance associated with the shot noise is first measured at various input powers by splitting the probe at a far off-resonance frequency (to avoid the rubidium resonances) 50:50 and directing it into the two arms of the balanced detector. Fig. 8 shows a plot of the intensity noise variance as a function of input power, with the linear slope indicating a shot-noise limited measurement. 

\begin{figure}[htbp]
\centering \includegraphics[width=10 cm, height=  8 cm]{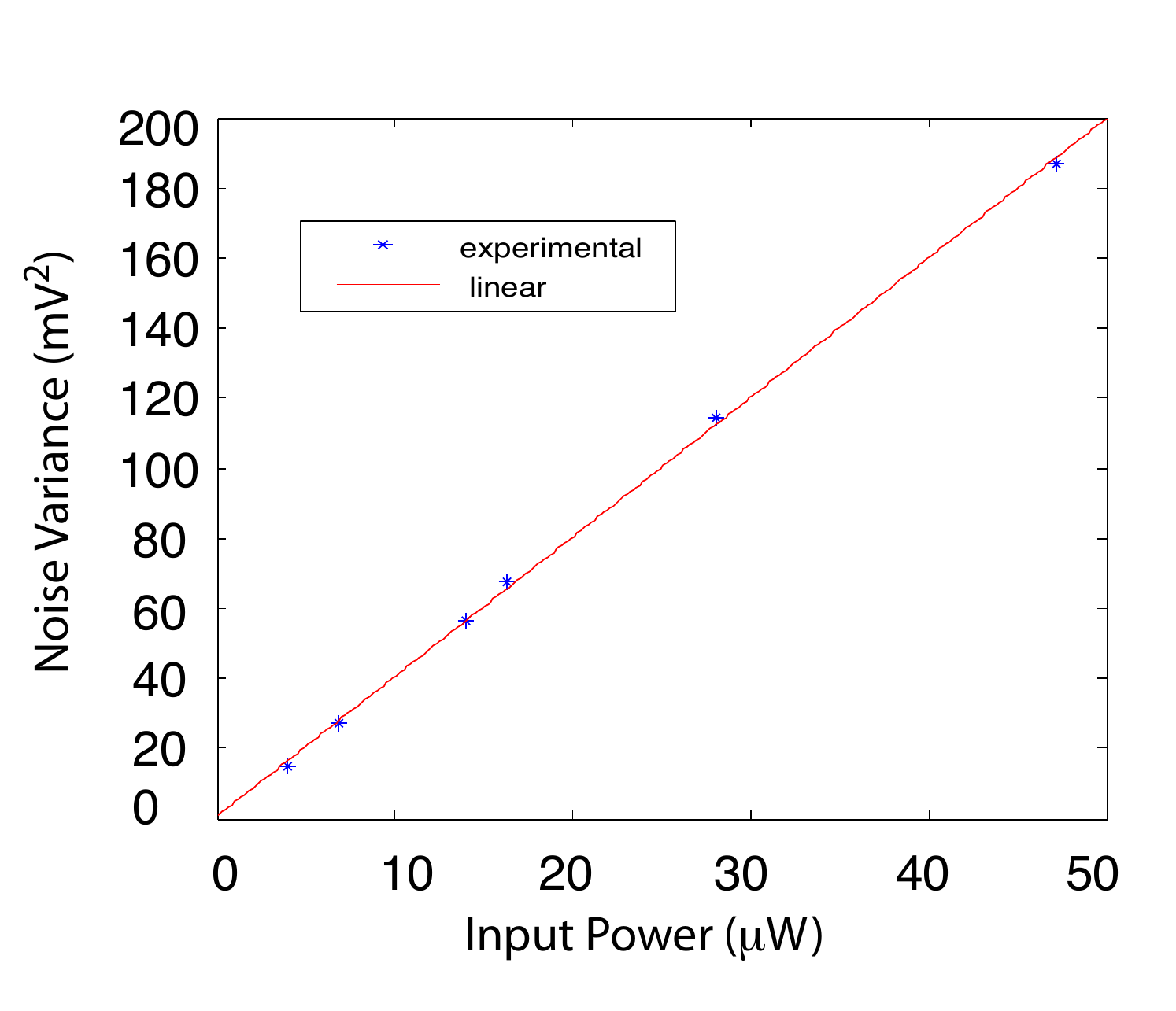}
\caption{The shot-noise quantum limit as a function of input power; the linear behaviour indicates shot-noise statistics.  }
\end{figure}

Next, with one laser and the two-photon detunings optimized for maximum gain, we record the noise variance of the probe-conjugate intensity difference after equilibrating the two arms by attenuating the probe channel. The fluctuations of the amplified output are first minimized by carefully equilibrating the probe-conjugate channels, and also, as we observed, by adjusting the two-photon detuning to about 20 MHz (as predicted by the theory). The integrated intensity difference is then recorded through the data acquisition card for both the shot-noise limit and for the probe-conjugate intensity difference (Fig. 9). 

\begin{figure}[htbp]
\centering \includegraphics[width=10 cm, height=  8  cm]{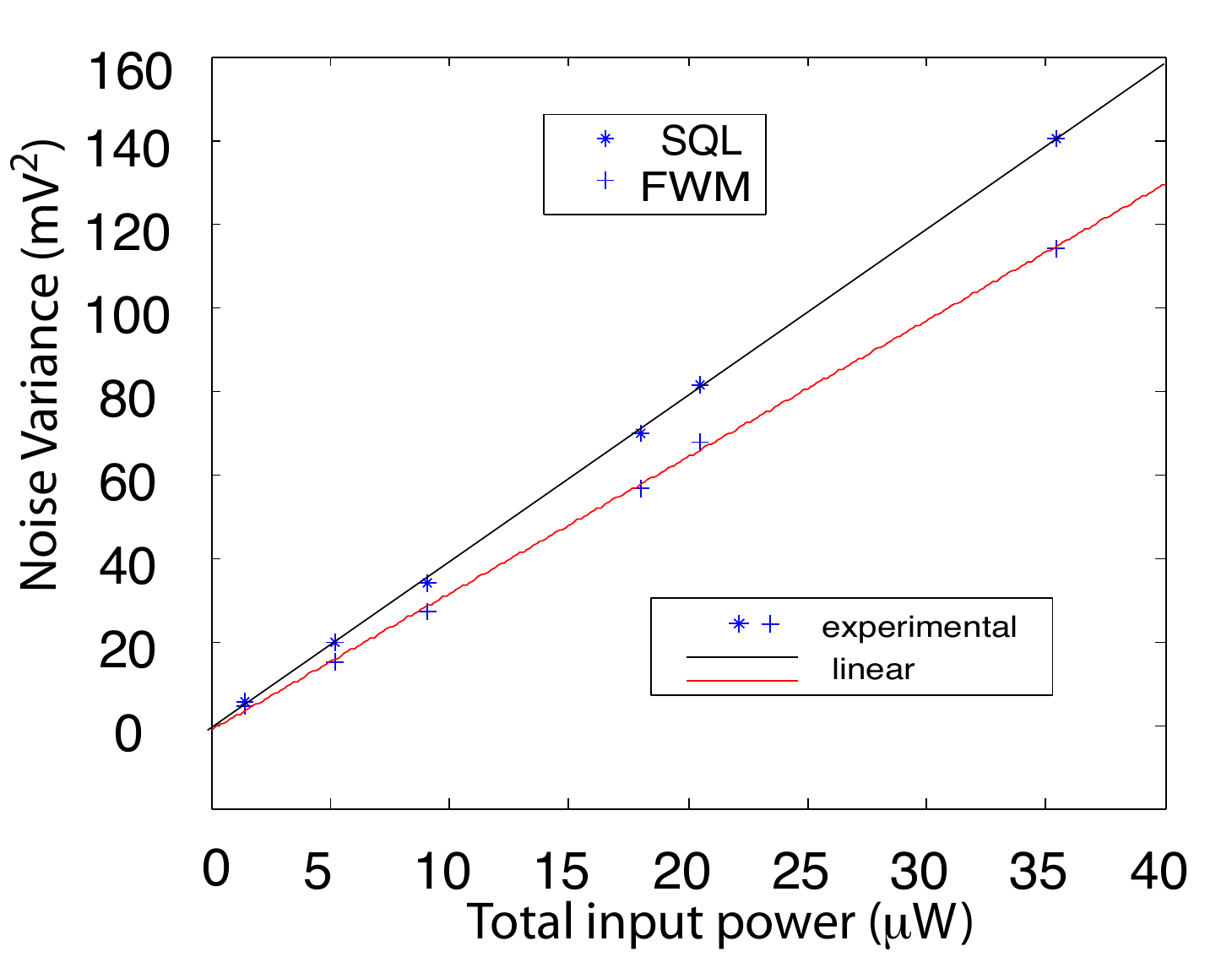}
\caption{Relative intensity noise for a 50/50 split probe (*) and probe-conjugate under four-wave mixing conditions (+) as a function of the total input power on the detectors. The ratio between the two variance curves at the same power yields the squeezing level.  }
\end{figure}

The results show relative intensity squeezing amounting to -0.96 dB, or -1.34 dB when corrected for the losses on the mirrors, lenses and when taking into account the estimated quantum efficiency of the detectors. Several factors contribute to the reduced levels measured as compared to the results of earlier experiments \cite{lett1, coudreau2} via frequency-resolved detection. First,  as the measurement is done in the time domain, the total noise is 'integrated' over the entire detection bandwidth (8 MHz), as shown in the theoretical section, which requires  us to use a two-photon detuning that best optimizes the pulsed squeezing rather than the single-frequency noise level and which consequently leads to a lower level of squeezing. Second, as the input probe is not a shot-noise limited state, but rather carries  excess noise of about 2 dB due to the electro-optic modulation, this may degrade the squeezing \cite{coudreau2} and explain the slightly lower degree of squeezing than that predicted theoretically (-1.34 vs. -1.6 dB). 

However, whereas earlier experiments showed the zero-bandwidth single frequency noise level, our results confirm the presence of squeezed light via the full photon statistics of the pulsed light. In other words, in a phase-sensitive homodyne measurement (rather than a relative intensity measurement), time-domain detection allows for a straight-forward measurement of the quadrature variances of the squeezed light. Moreover, in the context of information transfer and encoding, generating squeezed pulsed light (and temporally detecting it) allows for envisioning future experiments that harness the reduced noise levels for optical communications. In the context of quantum communications, squeezed pulsed light and its time-domain detection are a basic building block for several protocols for the generation of non-Gaussian states of light \cite{wenger} as well as for the encoding and secure transmission of quantum information \cite{cerf}. 

\section{Conclusion}
In conclusion, we have numerically shown the possibility of generating,  via the process of  nearly-degenerate off-resonance four-wave mixing in a hot rubidium vapor, nanosecond pulsed relative-intensity squeezed light. We have also measured the degree of squeezing experimentally by employing  time-domain balanced detection, leading to  -0.96 dB (-1.34 dB corrected) of relative-intensity squeezed 50 ns pulses at 1 MHz repetition rate. Our proof-of-principle experiment opens the door to future rubidium-based quantum memory demonstrations that require MHz repetition-rate rubidium-compatible squeezed light, and are a vital component in future quantum repeater protocols.  Moreover, by employing  time-domain detection of squeezed light, this experiment paves the way towards the de-Gaussification of the squeezed pulses and the generation of atom-compatible non-classical states that will be an important ingredient in atom-based quantum information processing.

\section{Acknowledgements}
I. H. A. would like to thank P. Londero and V. Venkatraman for helpful discussions. This work was supported by the EU grant COMPAS. I. H. A.  gratefully acknowledges the R\'egion Ile-de-France for a postdoctoral grant in the framework of the C'Nano IdF program. Q.G. and T.C. acknowledge fruitful discussions with Romain Dubessy, Ennio Arimondo,
Luca Guidoni, JeanÐPierre Likforman and Samuel Guibal.

\end{document}